\documentclass[aps,pra,twocolumn,showpacs]{revtex4}
\usepackage{eurosym}
\usepackage{amsfonts}
\usepackage{amsmath}
\usepackage{amsopn}
\usepackage{amssymb}
\usepackage{graphicx}
\usepackage{graphics}
\usepackage{color}
\usepackage{verbatim}

\begin{document}

\title{Nonlinear modes {in binary bosonic condensates} with the
pseudo-spin-orbital coupling}
\author{D. A. Zezyulin$^{1}$, R. Driben$^{2,3}$, V. V Konotop$^{1}$, and B.
A. Malomed$^{2}$}
\affiliation{$^{1}$Centro de F\'isica Te\'{o}rica e Computacional and Departamento de
F\'isica, Faculdade de Ci\^encias, Universidade de Lisboa, Avenida Professor
Gama Pinto 2, Lisboa 1649-003, Portugal \\
$^{2}$Department of Physical Electronics, School of Electrical Engineering,
Faculty of Engineering, Tel Aviv University, Tel Aviv 69978, Israel \\
$^{3}$Department of Physics, University of Paderborn, Warburger Str. 100,
D-33098 Paderborn, Germany}
\date{\today}

\begin{abstract}
We consider a binary Bose-Einstein condensate (BEC) with nonlinear repulsive
interactions and linear spin-orbit (SO) and Zeeman-splitting couplings. In
the presence of the trapping {harmonic-oscillator (HO)} potential, we report
the existence of even, odd, and asymmetric spatial modes. They feature
alternating domains with opposite directions of the pseudo-spin, i.e.,
anti-ferromagnetic structures, {which is explained by the interplay of the
linear couplings, HO confinement, and repulsive self-interaction. }The
number of the domains is determined by the strength of the SO coupling. The
modes are constructed analytically in the weakly nonlinear system. {The
dynamical stability of the modes is investigated by means of the
Bogoliubov--de Gennes equations and direct simulations. A notable result is
that the multi-domain-wall (DW) structures are stable, alternating between
odd and even shapes, while the simplest single-DW structure is\emph{\
unstable}.} Thus, the system features a transition to the \emph{complex
ground states} under the action of the SO coupling. The addition of the
Zeeman splitting transforms the odd modes into asymmetric ones via
spontaneous symmetry breaking. {The results suggest possibilities for
switching the binary {system} between states with opposite (pseudo)
magnetization by external fields{, and realization of similar stable states
and dynamical effects in solid-state and nonlinear-optical settings emulated
by the SO-coupled BECs}}.
\end{abstract}

\maketitle

\section{Introduction}

An important application of Bose-Einstein condensates
(BECs) is their use for emulating a variety of fundamental effects {%
originating in the realm of} condensed-matter physics, in a form which is
much easier to handle in {atomic} gases \cite{emulator}. Much attention was
recently attracted to the implementation of linear couplings between two
components of a binary BEC, which {emulate} the spin-orbit (SO) interactions
in solid-state settings, with the respective spinor order parameter {mapped
into} the two-component wave functions of mixtures~of different states of
the same atomic species \cite{Nature}. Besides the fundamental interest
concerning the dynamics of spinor BECs, the system represents a testbed for
the study of artificial gauge fields, which is another topic of a {rapidly
growing} interest~\cite{review}. To a great extent, the {possibility of}
emulating {condensed-matter effects in} ultra-cold gases was brought to the
focus of the current research by experiments~\cite{Abelian} demonstrating {%
the action of Abelian and non-Abelian} synthetic gauge fields in BEC~\cite%
{Ab-non-Ab}.

In addition to simulating {condensed-matter phenomena,} the studies of
SO-coupled BECs reveal new {matter-wave} effects, {produced by} the
interplay of the SO coupling and the {mean-field} nonlinearity induced by
atomic collisions. Such effects include sophisticated vortical \cite{vortex}
and monopole \cite{monopole} structures, multi-domain patterns \cite{Wang},
and patterns produced by long-range interactions~\cite{Santos}, tricritical
points \cite{Pitaevskii}, skyrmions \cite{skyrmions}, solitons~ {\cite%
{AFKP,latest}, interaction with optical lattices \cite{Liu}, etc.} Note that
the system of coupled Gross-Pitaevskii equations (GPEs) derived in~Ref. \cite%
{AFKP} is tantamount to that describing the co-propagation of two
polarizations of light in twisted nonlinear optical fibers~\cite{old}, hence
an additional link between completely different physical settings is the
possibility to emulate {birefringent optical fibers} by the SO-coupled
condensates, and \textit{vice versa}. Furthermore, a recent report on the
optical implementation of a model simulating massive Dirac fermions \cite%
{Dirac_optics} indicates the further relevance of results reported below to
nonlinear guided-wave optics.

The variety of expected effects in SO-coupled BEC may be greatly
enriched when three (or more) atomic levels are employed to define
relevant components and interactions between them \cite{review}. In
particular, \textit{tripod} atomic schemes give rise to coupled GPEs
featuring various nonlinear terms which do not amount to the usual
self-phase and cross-phase modulation~\cite{tripod}. Such systems
open possibilities for nonlinear control of SO-coupled BECs by means
of resonant laser illumination.

While spinor BECs typically give rise to two phases with {opposite
polarizations of }pseudo-spins, and to mixtures of such phases~\cite%
{Nature,Wang,Pitaevskii}, not every form of the SO coupling
results in energy splitting between the phases. For such a
situation, the role of the nonlinearity is crucially important, as
it lifts the degeneracy and induces the phase separation, cf.
Ref.~\cite{Barcelona}.

In this work we study nonlinear modes originating from degenerate linear
eigenstates in the SO-coupled binary BEC loaded into a {harmonic-oscillator
(HO)} trap, finding a full set of such nonlinear states. We produce basic
spatial patterns which exist in the case when the Rashba~\cite{Rashba} and
Dresselhaus~\cite{Dresselhaus} couplings have equal strengths. The patterns
are built of alternating domains with opposite directions of the
pseudo-spin, with each state characterized by zero or nonzero total
(pseudo-) magnetization. The number of domains is determined by the strength
of the SO coupling. The most essential new results concern the stability of
the competing states, which is determined by the total magnetization. In
contrast to previously studied settings, we demonstrate that the
multi-domain-wall (DW) patterns may be \emph{stable} while the single DW
\emph{is not}, hence the system's ground state shifts to the complex
patterns. We also report a possibility of the dynamical switching between
different stable patterns.

\section{{The model}}
We consider the two-component effectively
one-dimensional (1D) BEC described by spinor ${\mathbf{\Psi }}=\mathrm{col}%
\{\psi _{1},\psi _{2}\}$, whose components $\psi _{1}$ and $\psi _{2}$
represent pseudo-spin components $|\!\!\uparrow \rangle $ and $%
|\!\!\downarrow \rangle $. The dynamics of the system is governed by the
Hamiltonian, $H=H_{0}+H_{\mathrm{int}}$, where, in scaled units, $%
H_{0}=\int_{-\infty }^{+\infty }\mathbf{\Psi }^{\dag }\mathcal{H}\mathbf{%
\Psi }dx$, $\mathcal{H}=(1/2)(-\partial _{x}^{2}+x^{2}+\Omega \sigma
_{3})+i\kappa \sigma _{1}\partial _{x}$, with {Pauli matrices }$\sigma _{3,1}
${, and }$x^{2}/2$ is the axial trapping potential \cite{NJP}. Zeeman
splitting $\Omega $ is induced via a constant magnetic field {acting along
the} $z$-axis, while the SO coupling, {accounted by} coefficient $\kappa $,
results from {a combined effect of the Rashba and Dresselhaus couplings},
and is determined by {intensities and wavelengths} of laser beams {which
couple the relevant atomic levels.} The interaction terms for the underlying
{energy-level scheme }are represented by $H_{\mathrm{int}}=\!\int_{-\infty
}^{+\infty }\!\left[ (g_{1}/2)\left( |\psi _{1}|^{4}+|\psi _{2}|^{4}\right)
+g|\psi _{1}|^{2}|\psi _{2}|^{2}\right] dx$ ~\cite{tripod}. The derivation
of the 1D model from the full 3D one follows the scenario of introducing a
tight HO potential of the transverse confinement and factorizing the wave
functions into the ground state of that potential and a free longitudinal
wave function, which has been elaborated in detail \cite{1D}. The structure
of the SO-coupled system does not imply any problem in following this
scenario, provided that it leads to the usual cubic nonlinearity. The
situation may be different for relatively dense BEC, when the resulting 1D
model features deviations from the cubic interactions (for the binary
condensate without the SO coupling this situation was elaborated in Ref.
\cite{Luca}), which may be a subject for separate analysis~\cite{Luca1}.

The 1D Hamiltonian gives rise to coupled GPEs,%
\begin{gather}
i\partial _{t}\psi _{j}=-\frac{1}{2}\partial _{x}^{2}\psi _{j}+\frac{x^{2}}{2%
}\psi _{j}+i\kappa \partial _{x}\psi _{3-j}-(-1)^{j}\frac{\Omega }{2}\psi
_{j}  \notag \\
+\left( g_{1}|\psi _{j}|^{2}+g|\psi _{3-j}|^{2}\right) \psi _{j},\quad j=1,2,
\label{psi}
\end{gather}%
which conserve the total number of atoms, $N=N_{1}+N_{2}$, with $%
N_{1,2}=\int_{-\infty }^{+\infty }\left\vert \psi _{1,2}\right\vert ^{2}dx$.
Below we consider the generic situation ($g\neq g_{1}$), and keep $N$ as a
free parameter, fixing positive coefficients $g_{1}$ and $g$, which account
for the intra- and inter-species repulsive interactions, respectively.
Stationary modes of Eq.~(\ref{psi}) with chemical potential $\mu $
correspond to $\mathbf{\Psi }(x,t)=e^{-i\mu t}\mathbf{\Phi }(x)$, where
spinor $\mathbf{\Phi }=\mathrm{col}\{\phi _{1},\phi _{2}\}$ obeys the system
\begin{gather}
\mu \phi _{j}=-\frac{1}{2}\phi _{j}^{\prime \prime }+\frac{x^{2}}{2}\phi
_{j}+i\kappa \phi _{3-j}^{\prime }-(-1)^{j}\frac{\Omega }{2}\phi _{j}  \notag
\\
+\left( g_{1}|\phi _{j}|^{2}+g|\phi _{3-j}|^{2}\right) \phi _{j},
\label{phi}
\end{gather}%
with $\phi _{j}^{\prime }\equiv $ $d\phi _{j}/dx$. In the free space
(no trapping potential) and in the absence of the linear coupling,
{the commonly known} condition for the immiscibility of the binary
condensate is $g>g_{1}>0 $ \cite{miscibility,books}. The trapping
potential and Zeeman splitting shift the transition to the
miscibility from $g=g_{1}$ to larger values of $g $ \cite{we}. Note
that, unlike the equations considered in~Ref. \cite{AFKP},
model~(\ref{psi}) is specific to the SO-coupled system, as it did
not occur previously in fiber optics. In principle, there is a
chance to implement this system in nonlinear optics too, but in a
different context, using settings such as the recently reported Dirac model~\cite%
{Dirac_optics}, which is obtained from Eq. (\ref{phi}) by neglecting
the kinetic-energy terms.

\section{{Zero Zeeman splitting}} 
\begin{figure}[tbp]
\centering
\includegraphics[width=0.9\columnwidth]{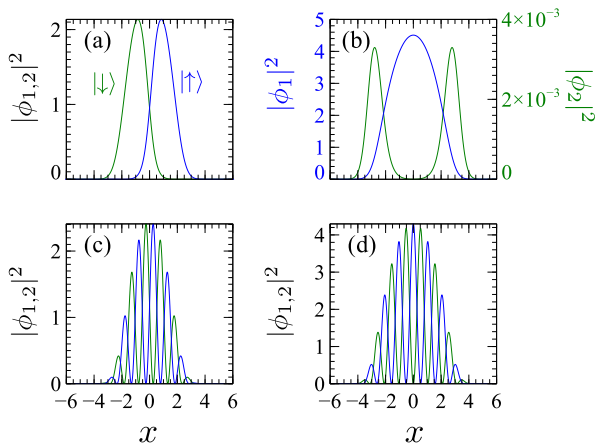}
\caption{(Color online) Generic examples of nonlinear modes bifurcating from
the lowest linear eigenstate with $n=0$. Left [right] panels display odd
modes for $N=8$ [even modes with even $\protect\phi _{1}(x)$ and odd $%
\protect\phi _{2}(x)$ for $N=18$]. Top [bottom] rows correspond to $\protect%
\kappa =0.1$ [$\protect\kappa =3$]. Modes (b)-(d) are stable, while mode~(a)
is unstable. Other parameters are $g_{1}=g/2=1$, $\Omega =0$. Different
vertical scales for $\protect\phi _{1}(x)$ and $\protect\phi _{2}(x)$ in
panel (b) indicate that $|\protect\phi _{1}(x)|\gg |\protect\phi _{2}(x)|$.}
\label{fig:n=0}
\end{figure}

At $\Omega =0$, Eqs.~(\ref{psi}) give
rise to an evident solution, $\psi _{1}=\psi _{2}=\exp \left( i\kappa
x+i\kappa ^{2}t/2\right) \chi \left( x,t\right) $, with function $\chi $
obeying the standard GPE: $i\partial _{t}\chi =\left[ -(1/2)\partial
_{x}^{2}+x^{2}/2+\left( g_{1}+g\right) |\chi |^{2}\right] \chi $. Below, we
refer to this solution, characterized by nonzero superfluid velocity, as a
\textit{current state}. At the same time, the symmetry of system (\ref{phi})
with $\Omega =0$ admits solutions of two other types: an \textit{odd mode},
with {$\phi _{2}(x)\equiv i\phi _{1}(-x)$}, where $\phi _{1}(x)$ and $\phi
_{2}(x)$ are purely real and imaginary functions, respectively; and an
\textit{even mode}, with {$\phi _{1}(x)$ and $\phi _{2}(x)$ having opposite
parities, e.g., $\phi _{1}(x)$ is even and real, while $\phi _{2}(x)$ is odd
and imaginary, see examples in Fig.~\ref{fig:n=0}. Odd and even modes
correspond to symmetric and anti-symmetric distributions of the
pseudo-magnetization density, $\mathcal{M}=\left( |\phi _{1}(x)|^{2}-|\phi
_{2}(x)|^{2}\right) /N$, whose integral }$M=\int_{-\infty }^{+\infty }%
\mathcal{M}dx=(N_{1}-N_{2})/N$ defines the total magnetization.

Possible types of nonlinear modes, which constitute one-parameter families
of solutions, can be identified by the analysis of bifurcations of the
families from the linear limit, $N\rightarrow 0$ (or, alternatively, $%
g_{1}=g=0$), which leads to the eigenvalue problem, $\mathcal{H}\tilde{%
\mathbf{\Phi }}=\tilde{\mu}\tilde{\mathbf{\Phi }}$ (hereafter, the tilde
stands for the linear limit).
It is easy to see that the spectrum of this separable system is
double-degenerate: $\tilde{\mu}_{n}=n+(1-\kappa ^{2})/2$, $n=0,1,2,...$
Eigenvectors, $\tilde{\mathbf{\Phi }}_{n}\equiv \mathrm{col}\{\tilde{\phi}%
_{1,n},\tilde{\phi}_{2,n}\}$, can be chosen as arbitrary superpositions of
two mutually orthogonal ($\langle \tilde{\mathbf{\Phi }}_{n,+},\tilde{%
\mathbf{\Phi }}_{n,-}\rangle =0$~\cite{inner}) spinors: $\tilde{\mathbf{\Phi
}}_{n}=C_{+}\tilde{\mathbf{\Phi }}_{n,+}+C_{-}\tilde{\mathbf{\Phi }}_{n,-}$,
where ${\tilde{\mathbf{\Phi }}}_{n,\pm }=e^{\pm i\kappa
x}e^{-x^{2}/2}H_{n}(x)\text{col}\left\{ 1,\pm 1\right\} $, and $H_{n}(x)$ is
the Hermite polynomial.

{However, an arbitrary set} of constants $C_{\pm }$ in $\tilde{\mathbf{\Phi }%
}_{n}$ {does not correspond} to nonlinear eigenmodes. To determine the
respective constraints, which implies \emph{lifting the degeneracy} by the
repulsive nonlinearity, we introduce expansions $\mu =\tilde{\mu}%
_{n}+\epsilon ^{2}\mu _{n}^{(2)}+\ldots $ and $\mathbf{\Phi }=\epsilon
\tilde{\mathbf{\Phi }}_{n}+\epsilon ^{3}\mathbf{\Phi }_{n}^{(3)}+\ldots ,$
which satisfy Eqs.~(\ref{phi}) with $\Omega =0$ at the leading order with
respect to $\epsilon $, which characterizes the strength of the
nonlinearity. At order $\epsilon ^{3}$, one has $(\mathcal{H}-\tilde{\mu}%
_{n}){\mathbf{\Phi }}_{n}^{(3)}=-\mathbf{F}_{n}+\mu _{n}^{(2)}\tilde{\mathbf{%
\Phi }}_{n},$ with
\begin{equation}
\mathbf{F}_{n}=\left( \!%
\begin{array}{c}
(g_{1}|\tilde{\phi}_{1,n}|^{2}+g|\tilde{\phi}_{2,n}|^{2})\tilde{\phi}_{1,n}
\\
(g_{1}|\tilde{\phi}_{2,n}|^{2}+g|\tilde{\phi}_{1,n}|^{2})\tilde{\phi}_{2,n}%
\end{array}%
\!\right) .  \label{F}
\end{equation}%
The solvability condition for ${\mathbf{\Phi }}_{n}^{(3)}$ results in two
equations, $\mu _{n}^{(2)}\langle \tilde{\mathbf{\Phi }}_{n,\pm },\tilde{%
\mathbf{\Phi }}_{n}\rangle =\langle \tilde{\mathbf{\Phi }}_{n,\pm },\mathbf{F%
}_{n}\rangle$, where constant $C_\pm$ are the unknowns. Focusing on
the lowest linear eigenstate (the one  which corresponds to $n=0$),
it is straightforward to find solutions of the latter equations. One
solution has $C_{-}=0$ and $C_{+}\neq 0$, the respective linear mode
being $\tilde{\phi}_{1,n}(x)=\tilde{\phi}_{2,n}(x)$. In the
nonlinear regime, it keeps the same structure, $\phi _{1}(x)\equiv
\phi _{2}(x)$,
representing the current state, as defined above. Another solution has $%
C_{+}=\pm iC_{-}$. In the linear limit, the associated eigenmode is $\tilde{%
\mathbf{\Phi }}_{0}=e^{-x^{2}/2}\mathrm{\,col}\left\{ \cos (\pi /4+\kappa
x),i\cos (\pi /4-\kappa x)\right\} \!$, which extend into the
above-mentioned \emph{odd} nonlinear mode, with $\phi _{2}(x)\equiv i\phi
_{1}(-x)$. %
{Finally,} the third solution, with $C_{+}=\pm C_{-}$ and $\tilde{\mathbf{%
\Phi }}_{0}=e^{-x^{2}/2}\,\mathrm{col}\left\{ \cos (\kappa x),i\sin (\kappa
x)\right\} $, represents \emph{even }modes in the nonlinear regime.
Importantly, \emph{no other solutions} exists in the weakly nonlinear system.

Thus we have identified three families of nonlinear modes
bifurcating from the lowest eigenstate $n=0$.
Numerically found examples of the nonlinear modes from these families are shown in Fig.~\ref%
{fig:n=0}. While, {as said above, one should expect the immiscibility} due
to the repulsive interactions, {most modes} feature the {striped}
antiferromagnetic structure, represented by alternating domains of states $%
|\!\!\uparrow \rangle $ and $|\!\!\downarrow \rangle $, rather than
two large domains {separated by a single (DW,} which is the case in
the conventional binary BEC~\cite{we,DWs}. {It is seen that the}
number of the domains increases with the {SO coupling strength}, the
single DW being present only in panel {\ref{fig:n=0}(a)}. This
patterning is precisely explained by the spatially periodic factors
in the above analytical solutions. In qualitative terms, the
interplay of the {HO trap with the SO coupling gives rise to scale}
$l\sim 1/\kappa $ that determines the periodicity of the striped
patterns in Fig. \ref{fig:n=0} (the analytical solutions yield
$l=\pi /\kappa $). It is relevant to mention that multi-domain
patterns were also reported in~Ref. \cite{Matuszewski} in the spinor
(three-component) BEC under the action of external magnetic field
(without the SO coupling). An essential difference of our system is
that the transition between different numbers of domains is
controlled by the intrinsic strength of the SO coupling, rather than
by an additionally introduced magnetic field. For the SO-coupled
system with spin $2$, two-dimensional multi-domain patterns were
reported in Ref. \cite{S=2}, but the stability of those patterns
(which is the main subject of the present analysis) was not studied.

\begin{figure}
\includegraphics[width=0.9\columnwidth]{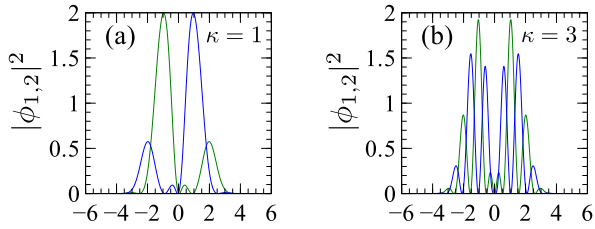}
\caption{(Color online) Examples of stable nonlinear modes
bifurcating from the first excited ($n=1$) linear eigenstate (rather
than from the ground state with $n=0$, cf. Fig. \ref{fig:n=0}.
Panels (a) and (b) show, severally, odd and even modes for
$g_{1}=g/2=1$ and $\Omega =0$. } \label{fig:n=1}
\end{figure}

A similar analysis has been performed for higher-order nonlinear
modes, i.e. for ones stemming from excited states of the linear
system ($n\geq 1$, see above). They also feature the multi-DW
structure of the odd and even types,  with the number of domains
increasing with the strength of the SO coupling, see the examples
displayed in Fig.~\ref{fig:n=1}.

\begin{figure}[tbp]
\includegraphics[width=0.9\columnwidth]{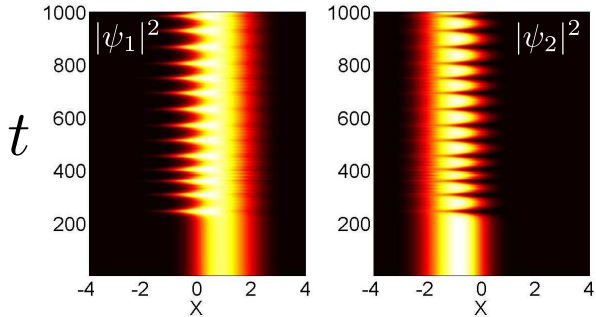}
\caption{(Color online) The evolution of the unstable  nonlinear
mode shown in Fig.~\ref{fig:n=0}~(a). Under the action of the
instability, the mode spontaneously transforms into a persistent
oscillatory state.} \label{fig:inst}
\end{figure}

In view of the coexistence of many multi-DW patterns, their
stability is a crucial issue. Here we are meaning the experimentally
relevant dynamical stability, determined by the Bogoliubov-de Gennes
(BdG) equations derived from Eqs.~(\ref{psi}), rather than the
thermodynamic stability. The spectrum of the BdG equations was found
by means of standard methods for the solution of the corresponding
eigenvalue problem \cite{books}, and the results were verified by
direct simulations of the perturbed evolution of the modes. A
surprising conclusion is that the simplest pattern with the single
DW, shown in Fig.~\ref{fig:n=0}(a), is \emph{unstable}, on the
contrary to the situation in the ordinary binary BEC \cite{we,DWs},
while the multi-DW patterns, displayed in panels (b-d), are
\emph{stable}. This finding may be explained by the fact that stable
are those structures which comply with the above-mentioned spatial
scale $l$ naturally selected by the system. In direct simulations of
the perturbed evolution, see Fig.~\ref{fig:inst}), the unstable
single-DW state spontaneously develops strong oscillations, in
agreement with the oscillatory instability predicted by the BdG
analysis. The resulting formation of robust dynamical modes in the
form of breathers is a physically relevant result too.

Furthermore, we have found that higher-order nonlinear modes, with
the number of domains increasing with strength of the SO coupling,
which bifurcate from the excited linear states, with $\mu
=\tilde{\mu}_{1,2,...}$ (see above), may also be stable in the
nonlinear system,  see examples of stable higher-order modes in
Fig.~\ref{fig:n=1}.

The full stability chart for the odd and even nonlinear modes originating
from $\mu =\tilde{\mu}_{0}$ is fairly complex, as shown in Fig.~\ref%
{fig:stab}. For sufficiently small values of $\kappa $ and in the linear limit $%
N\rightarrow 0$ 
only the even modes are stable. This is readily explained by the
fact that, at small $\kappa $, the system admits an obvious even
configuration with almost all atoms falling into a single state, see
Fig. \ref{fig:n=0}(b). For larger $\kappa $, the stability diagram
features a zebra-like structure, with alternating stability regions
of the even and odd modes (while {the aforementioned current states
are completely unstable, cf. Ref.~\cite{unstable-current})}. The
stability areas for the odd and even modes approximately (but not
precisely) complement each other, {which is a consequence of the
competition between coexisting nonlinear modes}
\begin{figure}[tbp]
\includegraphics[width=\columnwidth]{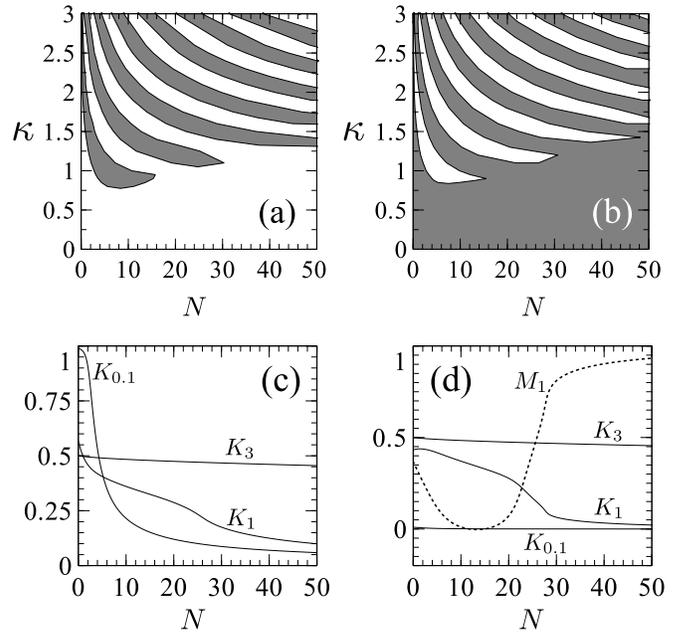}
\caption{(Color online) Stability domains (shaded) for odd (a) and
even (b) nonlinear modes bifurcating from the lowest linear
eigenstate, with $n=0$, in the system with the SO coupling in the
absence of the Zeeman splitting. (c) and (d): The overlap integral,
$K$, versus the total norm, $N$, for odd (c) and
even (d) modes. Curves labeled $K_{0.1,1,3}$ correspond to $\protect\kappa %
=0.1$, $1$, and $3$, respectively. The dashed curve in (d) shows the total
magnetic moment, $M$, as a function of $N$ for $\protect\kappa =1$. For $%
\protect\kappa =0.1$ and $\protect\kappa =3$, $M$ does not vary
significantly with $N$. Other parameters are $g_{1}=g/2=1$, $\Omega =0$.}
\label{fig:stab}
\end{figure}

These stability results are the most essential findings reported in this
work, as they demonstrate the previously unreported transition to \emph{%
complex ground states} in the system, under the action of the SO coupling.
In fact, this transition may be expected in diverse systems beyond the model
of the linearly-coupled binary BEC.

The stable patterns are naturally characterized by the total
pseudo-magnetization $M$ ($M=\pm 1$ means that the condensate is in a
single-domain state), and by the miscibility factor,
\begin{equation}
K\equiv 4\int_{-\infty }^{+\infty }\!\!|\phi _{1}|^{2}|\phi
_{2}|^{2}dx\left/ {\int_{-\infty }^{+\infty }\!\!(|\phi _{1}|^{2}+|\phi
_{2}|^{2})^{2}dx}\right. ,
\end{equation}%
which takes values $0\leq K\leq 1$, with $K=0$ and $K=1$ corresponding,
severally, to the completely immiscible or miscible state. Note that the
immiscibility, $K\rightarrow 0$, may be achieved not only if the phases are
spatially separated, but also if one phase disappears, and the condensate
falls into a single-domain state, as in the case of the even mode in Fig.~%
\ref{fig:n=0}~(b). In the linear { limit $N\to 0$}, one has $\tilde{M}=e^{-\kappa ^{2}}$
for the even eigenstates, and $\tilde{M}\equiv 0$ {for the current and odd
ones,} while $\tilde{K}=(1\mp e^{-2\kappa ^{2}})/2$ for the odd ($-$) and
even ($+$) states.

Figure~\ref{fig:stab}(c) shows that the nonlinearity enhances the
immiscibility of the odd modes, although the effect may be significantly
reduced by the coupling, as curve $K_{3}$ demonstrates. The effect of the
interplay between the nonlinearity and SO coupling is more sophisticated for
even modes, see Fig.~\ref{fig:stab}(d). When the SO coupling is sufficiently
weak (curve $K_{0.1}$) or strong (curve $K_{3}$), the nonlinear modes
preserve properties of their linear counterparts: at $\kappa =0.1$ the mode
is nearly fully polarized ($M_{0.1}\approx 1$, $K_{0.1}\ll 1$),
{while the mode} with $\kappa =3$ ($M_{3}\ll 1$, $K_{3}\approx 0.5$)
represents a partially miscible state, where both components are equally
represented, with $M$ close to zero.

The most interesting behavior is observed at intermediate values of the SO
coupling, {which corresponds to }$\kappa =1$ in panel (d). In this case, by
changing norm $N$ one draws the condensate into the immiscible state (curve $%
K_{1}$), and, at the same time, changes $M$ between zero and its extreme
value $+1$. Bearing in mind that the BEC nonlinearity may be controlled via
the Feshbach resonance \cite{books}, the observed behavior suggests a
possibility of switching between the different phases with the help of
external fields.

\section{ {Symmetry breaking due to the Zeeman splitting}}

\begin{figure}[tbp]
\includegraphics[width=\columnwidth]{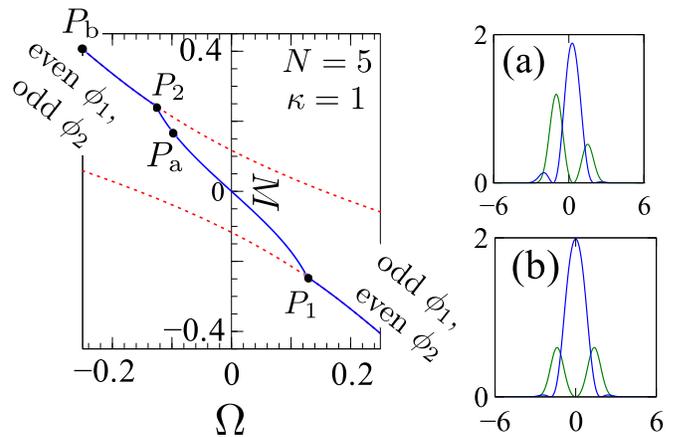}
\caption{(Color online) The bifurcation diagram (the left panel) for
$\protect\kappa =1$ and $N=5$, in the presence of the Zeeman
splitting, $\Omega $, the other parameters being as in
Fig.~\protect\ref{fig:stab}. Stable (unstable) modes correspond to
solid (broken) fragments of the curves. Panels (a) and (b) show
profiles of stable modes marked by points $P_{\mathrm{a,b}}$.}
\label{fig:dmu}
\end{figure}

 In the {
linear limit $N\to 0$}, the Zeeman splitting [$\Omega \neq 0$ in Eqs. (\ref{psi})] removes
the energy degeneracy. Now, the system does not admit current states and odd
{modes, while} stable even modes, characterized by even $\phi _{1}(x)$ and
odd $\phi _{2}(x)$ (or \textit{vice versa}), persist at $\Omega \neq 0$. The
evolution of the eigenstates under the action of the increasing Zeeman
splitting, $\Omega $, is illustrated by the bifurcation diagram in the plane
of $(M,\Omega )$, shown in Fig.~\ref{fig:dmu}. Starting with the point where
an odd (even) mode is stable (unstable) at $\Omega =0$, we observe a
transformation of the stable odd mode into stable asymmetric ones [see an
example in Fig.~\ref{fig:dmu}~(a)], {whose branch} approaches the branches
of the even modes. The instability of the even modes persists up to the
point where the branches of even and asymmetric modes merge (points $P_{1,2}$%
). At these points, the asymmetric modes disappear and the symmetric
even ones become stable [an example is shown in
Fig.~\ref{fig:dmu}(b)]. Thus, a pitchfork bifurcation occurs at
points $P_{1,2}$, {which is} a typical
example of the \emph{spontaneous symmetry breaking} (or restoration) \cite%
{breaking}.

\begin{figure}
\includegraphics[width=\columnwidth]{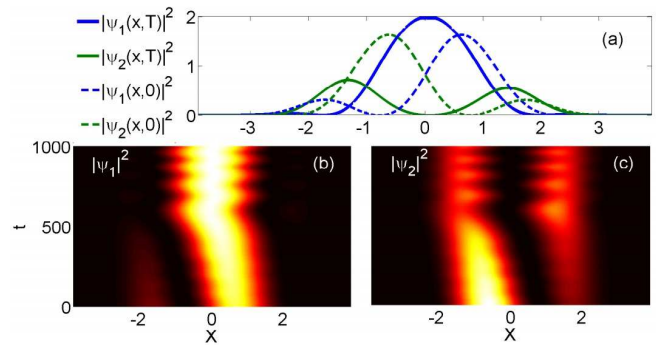}
\caption{(Color online) The evolution of a nonlinear mode subjected
to an adiabatic change of $\Omega $. Panel (a) shows the input and
output spatial profiles. Panels (b) and (c) display intensity plots.
The input at $t=0$ is a stable odd mode at $\Omega =0$, the other
parameters being as in Fig.~\protect\ref{fig:dmu}.
In the course of the simulation, $\Omega $ was adiabatically changed from $%
\Omega =0$ at $t=0$ to $\Omega =-0.25$ at $t=10^{3}$. As a result, the
condensate switches from the initial odd mode to an even one, as predicted
by the diagram in Fig. \protect\ref{fig:dmu}.}
\label{fig:dyn}
\end{figure}

The bifurcation diagram in Fig.~\ref{fig:dmu} suggests that, by varying the
Zeeman field {which induces } the Zeeman splitting, one can perform a
controllable switch between two opposite magnetizations in the SO-coupled
binary {system}. This suggestion is confirmed by simulations, as shown in
Fig.~\ref{fig:dyn}.

\section{Conclusion}

 We have reported the existence of even, odd, and
asymmetric nonlinear modes in the effectively 1D self-repulsive binary BEC
with the SO and Zeeman splitting, confined by the axial HO potential. {The
interplay between the coupling and the potential gives rise to the modes}
{featuring} alternating domains with opposite directions of the pseudo-spin,
which is explained analytically in the case of the weak nonlinearity. {%
Noteworthy findings are the stability of the multi-DW patterns, while the
one with the single DW is unstable, and the stability alternation between
the even and odd structures.} The implication of these results is the
transition from simple to complex ground states, driven by the SO coupling.
The inclusion of the Zeeman splitting results in the transformation of even
modes into asymmetric ones, which suggests a possibility of {controllable}
switching {\ between states }with opposite pseudo-magnetization. {These
effects, including the transition to the complex ground states, may be as
well expected in other physical setting emulated by the SO-coupled binary
condensates, such as solid-state media and bimodal guided-wave propagation
in optics.}

VVK and DAZ acknowledge support of the FCT (Portugal) grants
PEst-OE/FIS/UI0618/2011, and SFRH/BPD/64835/2009. The work of RD and BAM was
supported, in a part by the Binational (US-Israel) Science Foundation
through grant No. 2010239.


\begin{thebibliography}{99}
\bibitem{inner} The inner product used here is defined as $\langle \mathbf{a}%
,\mathbf{b}\rangle =\int_{-\infty }^{+\infty }[a_{1}^{\ast
}(x)b_{1}(x)+a_{2}^{\ast }(x)b_{2}(x)]\ dx$.
\end{thebibliography}
\end{document}